# ON SCALING FUNCTIONALITY IN URBAN FORM

Romulo Krafta
Federal University of Rio Grande do Sul, Brazil
krafta@ufrgs.br

ABSTRACT

Assuming that the ultimate purpose of the city is to provide support to human interaction and that opportunities to that social interaction are unevenly distributed across the urban fabric, this paper reports some attempts to describe such a distribution, as well as to infer the role of urban form in it. In order to do that, it is proposed, firstly, a method to describe urban form from its smallest components up to the different urban fabric patches, to the entire spatial system, and second, a model to represent social interaction as a process associated to the urban morphology. Both the spatial description and the analytical model are discussed through the examination of some results, obtained through simulation.

INTRODUCTION

Cities are places for social interaction (Batty, 2012, Bettencourt 2013, Jacobs, 1969). Social interaction occurs across cities in many different ways; it can vary according to group size, purpose, intensity, frequency, group composition and so one. It also can take place in different locations and settings, from the privacy of residential rooms to many collective gathering buildings to the open public space. All of them, regardless scale, type, location, etc., contribute to the social output of a city. In this sense, social output encompasses the so-called productive activities, as expected, as well as many others that invest in family, friendship, love, culture and civilization. All of them flourish in the city, some in concentrated ways, others scattered around, some daily, some occasionally, and depend upon the city to carry on.

The most recent and beautiful attempt to capture such a process in a unified urban model and to describe it quantitatively has been made by Bettencourt (2013); his model builds upon the notion that each individual has an actual interaction area, defined by the average distance he or she can cove in a day. The more mobile an individual is, the larger is its interaction, and consequently its social output. But area is not the only variable for interaction definition, as the presence of other people is essential, so Bettencourt considers an average density across the entire urban area, so that eventual more disperse areas will be compensated by others more dense. Mobility is, of course, defined by urban infrastructure, which puts urban actual form as crucial to social interaction and output.

Urban form is uneven and hierarchical, it is also occupied by so many different activities which contribute to its unevenness and hierarchy, some of those activities add to the basic urban form differentiation, some others can create their own differentiation. Moreover, mobility can vary widely within the urban fabric, all those evidences suggest that, despite the average described by Bettencourt's model being accurate, not only relevant differences in social interaction can be found, but also urban spaces, patches of urban fabric can perform better or worse, according the their morphologic characteristics. The possibility of achieving a more morphologic based representation of social interaction distribution suggests that not only our understanding of the phenomenon would improve but our ability to act upon key components of city spatial structure, aiming at performance improvement, could be better.

THEORY, MODEL DEFINITION

Assuming that the urban fabric is an aggregate of many components, it is relevant to identify what they are, and how they are put together, the smallest urban bits, the urban atom. The figure below tries to mirror the standard model of the atom, suggesting that the urban atom would have two types of matter particles – built forms BF and open public space PS, directly connected (A column in the figure). Encapsulated in those matter particles it is found other non-matter particles, called interaction-promoters IP (B column). These are force-carrier particles of various kinds; associated to BFs one could find all sorts of institutions (families, shops, service providers, industries, …), promoting specific interactions; associated to public spaces, other group of more ill-defined institutions, promoting less controlled interactions (C column).

In C, the figure suggests different types of social interactions, such as SI1, which could be strictly residential, SI3 & SI4 as two different types housed in a same built form, and SI2 as a more complex social interaction process involving the same people of SI1, 3 and 4, as well as other operators passing through the open public spaces. In this sense, the urban atom would entertain a same group of people (residents and eventual service users) in different types of social interactions, some more defined and controlled, indoors, and some less defined and controlled, outdoors. It becomes clear that some social interactions have trivial descriptions (residential, both isolated and grouped in condominiums), some are a matter of more careful consideration, as they do not have a previously known population (shops, offices, schools, hospitals), while public space is a mixture of local and passing by population.

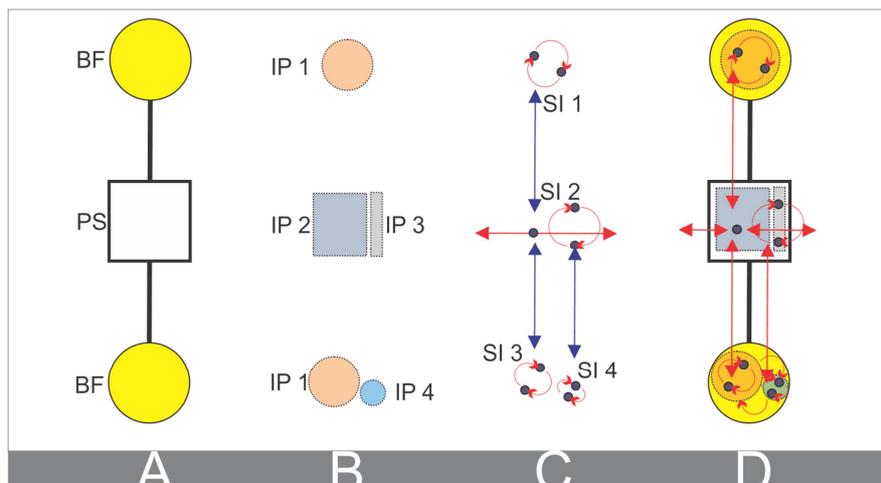

*Figure 1. An illustration of a urban atom: A: matter-particles built forms BF and open public space PS; B: interaction promoters IP of different types, each one corresponding to an social institution; C: social interaction associated to each matter-particle (red cycle arrows) and accumulation in the public space (blue arrows)*

Urban molecules of different sizes and shapes are easily derived from the atom, such as suggested in figure 2A. In the example, other particles are included, as open collective (although not public) spaces, such as front gardens, halls and other built spaces available to more than one interaction promoter IP. Molecules are, of course, joint together, forming the actual fabric, as in 2B, and the whole city.

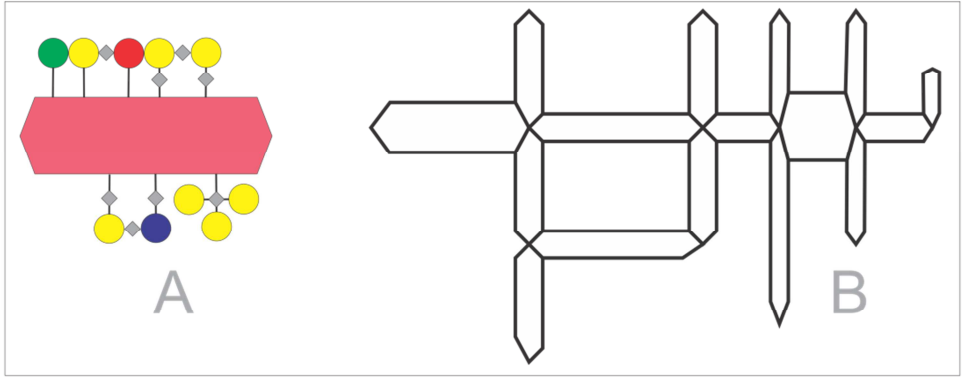

*Fig. 2: Aggregates of urban atoms; in A a urban 'molecule' made out of built forms BF (○○○○), collective open spaces (◊) and public open space (diamond); in B a second order aggregation of many interconnected molecules.*

Now, considering that every instance of spatial configuration is likely to house some kind of social interaction, the next step is to provide formal description to them. Social interaction is, of course, a function of the number of individuals present at a certain place, interaction with each other. The least a social interaction requires is two people, whereas social interaction could involve pairs, as well as any other amount of individuals. The larger the group, the higher the social interaction, in geometrical progression. Here social interaction will be taken by its root variable, population size P, relativized by a parameter K, which measures the specific social interaction's relevance (intensity, frequency, proportion to the whole social output). Social interaction is not cost-free, however, as individuals usually travel, sometimes striving to get to distant places, to put themselves in the right place at the right time.

$$SI = f(P, K, (-C)) \quad (1)$$

Starting from the smallest particle of the urban atom, the residential BF's social interaction population is the respective family, its relevance is probably high and the cost is nil, as the individuals involved in the actual social interaction are already there. In the equation 2, the expression $(P^I * C^i)$ implies that cost is per capita, meaning that in order to have interaction, $C^i$ should be <1. For residential BFs, Ci is zero. It is noteworthy that the opposite situation, in which cost is maximized, interaction could be null, or even negative. In this sense, interaction costs can prevent interaction, creating a sort of interaction passive. The possible values of K will be addressed later on.

$$SI(BF_{res}^i) = [P^I - (P^I * C^i)] * K^{resBF} \quad (2)$$

While residential BF's social interaction is trivial, non-residential ones are more complex, due to the fact that their actual population is not declared, as in residences, but made out of individuals that travel from many other places, in order to interact at the given BF. One possible way to proceed with its calculation is to consider all interaction promoters, that is, service providers of one type competing for all potential consumers; in this situation, every individual qualified as potential consumer is likely to choose the place of interaction based on least distance, submitted to size and complexity of the offered interaction places. For every service type, the whole pool of potential consumers will be split up among all places of provision of that service, in different proportions, according to spatial distribution of population and size/complexity of places of provision. A measure of that configuration, called *Convergence*, is in Krafta (1996), and for the purpose of this work, could be taken as a measure of attractiveness of every non-residential BF, helping to define each BF potential population.

Despite being attracted to a specific IP, individual will necessarily be located at different distances from it, meaning that each individual will bear a cost to travel from home to the considered IP; such a cost can be appropriated through a travel time, taken as the sum of each shortest path's spaces' time, measured by the quickest mean of transport, at worst traffic situation. Normalized, so that the shortest travel time being zero (the case of an individual living in the same space as the IP) and the longest being one (the most faraway place of living) each individual's actual travel time can be taken as its probability to carry out the travel&interaction; the chance of the individual living in the same space as the IP being one hundred per cent, whereas the most distant place being zero. In this sense, for each pair of spaces in which A is where the non-residential BF is located and B is where the residence of a potential consumer is located, the equation 2 applies, provided that population in B is resized by the convergence value of A, and the cost is the sum of travel times of all spaces linking A to B through the shortest path. The overall social interaction in the considered non-residential BF will be the sum of all pairs A-n provided by the system. K is, of course, very high.

$$SI(BF_{IPn}^i) = \sum_i^j [(P^j * Conv_i) - (P^j * Conv_i * C^{ij})] * K^{IPn} \qquad (3)$$

$$C^{ij} = \sum_j^p t^{jp} \qquad (4)$$

In the equation 3, Conv$^{IPn}$ (convergence) varies between 0-1, the same occurring to C$^i$, so that the BF population will always be a part of total potential consumers of IPn, and the total cost will be at most equal to total population. It is notable that total population of all IPs is necessarily bigger that the total population of the city, as most individuals will be potential consumers of more than one service, so that the whole numbers need a further normalization based on the actual city population.

Collective spaces are usually placed in between BFs, the private realm, and the open public space. Social interaction in there is performed by the same people living or using the BFs attached to the referred CS. Reaching the CS from each BF involves a cost which is not measured only by the corresponding (usually very small) elapsed travel time, but also the overcoming of an institutional barrier, a locked door of sorts. K is low.

$$SI(CS_{IPn}^i) = \sum_i^j [P^j - (P^j * C^{ij})] * K^{IPn} \qquad (5)$$

Open public spaces PS's social interaction has two basic components: locally it works just like a collective space, in which the population is the sum of all residents and other people attracted to it by the interaction promoters. Added to it there is the moving population, that is, people passing through. Strong references on co-presence, that is, the relationship between local people and individuals traveling across city spaces are in Hillier and Hanson (1984). Estimation of flows, based on morphology is in Krafta (1994), a measure of centrality derived from Freeman's betweenness centrality, to which two key changes were made. Betweenness centrality basically says that an entity C is central to a pair AB of other entities when it falls on the shortest path linking A to B. Overall centrality of C is measured by the number of times it falls on the shortest paths linking all possible pairs of entities of a system. Krafta argues that, in order to better adapt to spatial analysis, betweenness centrality ought to consider two more relationships: a) that centrality would be affected by the length of each shortest path, in which case a specific space's centrality would be higher in shorter shortest paths and lower in longer ones, and b) centrality would be affected by existing tension between the two spaces the referred space is in between. It is assumed that the contents of spaces can give more or less relevance to each pair, and, by extension more or less centrality to spaces in between.

To each person present at a public space corresponds a travel cost. Residents will have to cross one or two institutional borders, service users are already there, their travel cost already accounted for. Following the method previously adopted, it can be assumed that travel cost will range between 0 and 1, meaning that people moving from the most faraway place in the system will have cost 1, which in practice make its interaction null. Social interaction in public spaces is, therefore, represented by the following equation 6.

$$SI(PS^i) = \in \left[[P(BF_{res}^i) - C_{res}] + [P(BF_{IPn}^i) - C_{IPn}] + [P(flow) - C_{flow}]\right] * K^{PS} \quad (6)$$

TERRITORIAL DOMAIN

The method explained above suggests that travel time, or cost, is always maxim at the most faraway location, which is obvious, and equals 1, meaning that the whole system's population is likely to be included in the global interaction process, in the sense that even the most remote residential location will have a travel cost < 1 and therefore an interaction value >1. More realistic and interesting is to consider that this condition does not always hold, and thresholds should be considered, such as Marchetti's (1994) invariant – half an hour, or the observed city's average travel time. With thresholds on, every BF holding a non-residential Interaction promoter IP will have a territorial domain, constituted by all reachable spaces around it within the time limit. In this case, the normalized travel time value 1 won't be at the system's border, but on the threshold, and all spaces outside the territorial domain will be excluded from the considered BF interaction. This could be considered an interaction deficit, to be compared to the actual social interaction, resulting in an indicator of performance M*.

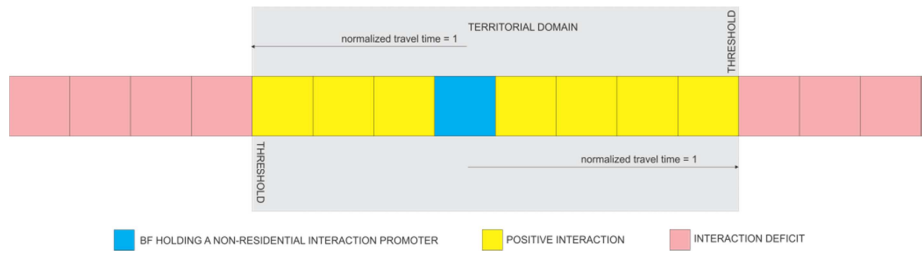

*Fig. 3: The notion of territorial domain in a one-dimensional city, defining the part of the system holding actual interaction in a particular space, as well as the part of the system excluded from it, therefore bearing a interaction deficit. The difference between positive interaction (yellow zone) and interaction potential unfulfilled (pink zone) is an indicator of performance M*.*

Establishing a threshold is not trivial, as both travel mode and urban form are involved. In general terms, urban form can be described through length, capacity and traffic density. The latter can easily be estimated by one of many accessibility measures available, working as a proxy for inverse speed and parking opportunity. However, such a description suits only one transportation mode, the private car, or public, surface, transport systems that use ordinary roads. For pedestrian travel mode, capacity is not relevant and speed is more uniform. Multi-modal travels are even more complex.

Two territorial domains seem to be relevant: the first related to purpose travels ending at the space being considered, and travels using that space as a link to elsewhere. For the latter, only pedestrian travel will count, as people passing through in cars or buses won't interact; for the former, all means should be considered. Territorial domain for pedestrians can be established once a distance/time limit is defined, using a simple accessibility algorithm. TD for people traveling to the considered space has two options: point-to-point car travel, and pedestrian-public transport-pedestrian one, provided that public transport runs on segregated lanes and are, therefore, free from the ordinary traffic congestion. In these cases, it will be necessary to proceed with time calculation, from length, capacity and traffic density, then to apply an algorithm which builds up a spanning tree from the space in question up to the threshold limit, calculating for every space visited, a population/cost factor.

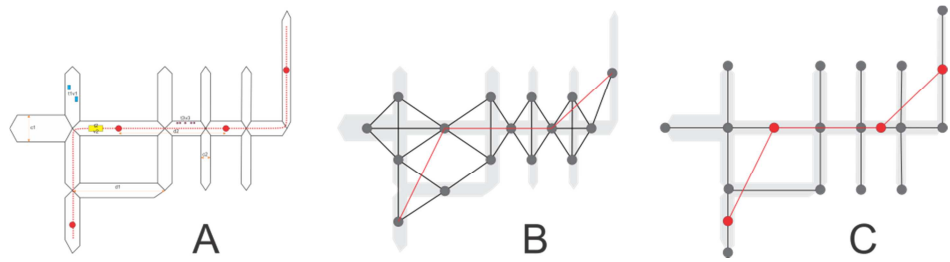

*Fig. 4: Spatial system (A) and graph representation of alternative space unit description: (B) space is described as street segments, in (C) corners and street ends.*

Space description is relevant here, figure 4 shows an idealized spatial system (A) and two graph-representations, taking different space descriptions; the graph (B) considers space units as street segments, whereas (C) takes space units as corners and street ends. It is easy to realize that B representation is more intuitive and probably handier for population and activity appropriation; on the other hand, C is better for distance/time description. Figure 4 shows also a possible public transport route (red dots and lines), running on segregated lane, so that its actual speed is different from the ordinary traffic.

Up to this point, it has been described all interaction that a space unit can generate by itself, both from its own population and from the visitors it can attract. However, nearby space units, particularly those bearing commercial activities, are known to benefit from that propinquity, case in which people visiting one is encouraged to visit, and therefore, interact in the others as well. To this extent, space units can work together, as a macro molecule (MM). Territorial domain is clearly of pedestrians, spatially defined by adjacency and increasing cost varying with distance. Equation 7 tries do capture this interaction

$$SI(MM_{IPn}^i) = \in \left[[P(PS_{IPn}^j) - C_{ij}]+\right] * K^{IPn} \qquad (7)$$

Finally, the K parameter should be addressed; it has been valued as 'high', 'low, 'very low', etc., suggesting that each type of interaction adds differently to the overall social output of a city and therefore K can be estimated. Indeed, it is easily seen that some interactions involve large number of people, compared to others that take just a couple of it or little more, some are very frequent, compared to others that occurs rarely, some are very well programmed and controlled, compared to others that are next to random. All of them are socially required and productive, each one in its own way, probably even those which look destructive. One possible way forward is to consider the relative density of social interaction over the population involved. On the one hand there is the residential instance, in which few people develop and maintain constant interaction of each family member with every other; on the other hand there is the public space, where a large population develop a few, eventual interaction. In the former, it takes few people to get high level of interaction, in the latter it takes large population to get low lever of interaction.

EARLY EXPERIMENTATION

In order to provide some early evidences on the effectiveness of the proposed models, some controlled experiments have been carried out. The city has been reduced to a one-dimensional string of sequentially connected cells, in which one half side is removed, as showed in figure 5. In this way, the cell on the far right is the centre. This simplified city is supposed to grow, from its most initial 2-cell embryo, up to 21 cells; in all its stages, the cells are filled with an Alonso style (1964) activity and population distribution, although at some point a secondary service centre is introduced, as explained later on. Total population in each stage is always 3X the number of residential cells.

Considering that built form constitution inside each cell is not specified (assumed as a single built form), residential interaction computation is trivial – equals the population of each cell, the same occurring with collective space's interaction (non-existent), and public space (again, interaction there equals the residence's one). A little more interesting is to observe how interaction in the centre occurs. Having no residential population of its own, the red cell attracts service users from all existing residential cells, who have to travel along the string. The simplified city implies that all cells are alike, meaning that infrastructure (roads) is equal; in this way, it is expected that traffic fluidity is maxim at the border and minimal at the centre (more congested). Traffic differential conditions at each cell have been taken as the inverse of accessibility, that is, at the centre, where accessibility is higher, traffic conditions is worse and vice versa. Length is equal for every cells, what makes the inverse of accessibility the only variable defining the travel cost from each residential cell to the CBD; this value has been initially normalized from zero (at the centre) up to one (at the border), meaning that all population is within the centre's range, although the cost of each one's actual interaction varies with distance. The computation of each residential cell's contribution to population at the centre is then made as a probability of each resident to take the trip; the ones living at the cell next to the centre is near 100%, the ones living at the most faraway is near 0%.

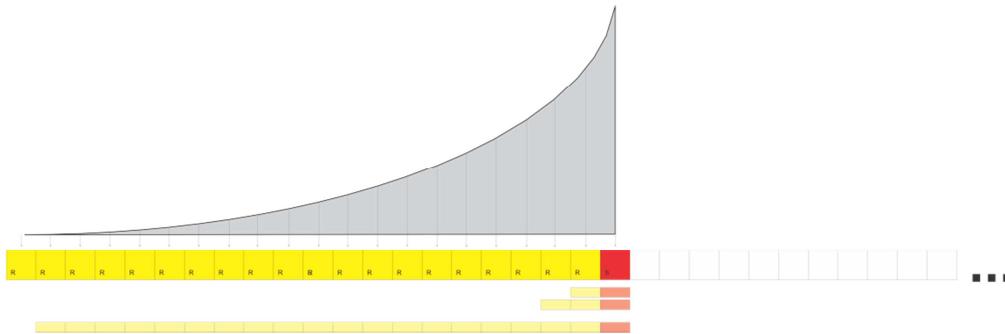

*Fig. 5: A One-dimensional half city made out of a string of 21 sequentially connected cells. It is supposed to grow from a 2-cells embryo up to 21 cells. Population distribution is as Alonso's general city form curve. Jobs and services are at the centre.*

Table 1 shows some results from a full 21 cell system, considered in three different ways: *a*) having a evenly distributed population (3 residents in each cell) and no traffic fluidity influence (only plain distance), *b*) population distributed according to Alonso's general curve and no traffic fluidity influence, and *c*) same as b, with traffic effect taken into account. Situation *b* produces the highest interaction output, whereas situation *c* the lowest. This is quite interesting, for it, first, do suggest why Alonso's monocentric, non-linear population distribution's city is so right, as it seems to be the most efficient interaction generating spatial configuration. It also suggests that it only works when the undermining effects of congestion is somehow minimized, otherwise it loses its density advantages. Table 1 also suggests that interaction taking place at the centre involves only a portion of the total population, varying from 45 to 75%. This however does not mean that 25 to 55% of it is excluded from it, but that not everybody will be interacting at a given time, although everyone will interact at some time.

*Table 1: Results for a 21 cell system, considered in three different situations: A) having a evenly distributed population and not suffering the effect of traffic congestion, B) having a non-linear distribution of population and not affected by traffic congestion, and C) same as b, with traffic effects.*

|                    | A    | B  | C  |
|-------------------:|:----:|:--:|:--:|
| Total population   | 60   | 60 | 60 |
| Total interaction  | 28,5 | 45 | 27 |
| % pop/interact     | 47,5 | 75 | 45 |

As a second step, the one-dimensional city has been examined while it grows from the 2-cell seed. It has been done through the evolution of the M* interaction performance indicator. M* is defined as the difference between actual produced interaction at a nominated space and the interaction deficit at the same place. Interaction deficit is understood as the unperformed interaction. In this sequence of experiments, a threshold point has been adopted. Threshold point defines the territorial domain for the considered centre, it is assumed that, either for preference or travel restrictions, interaction will only be possible for people located within the territorial domain, leaving anybody outside it excluded from it. Inverse accessibility, as well as travel times are calculated so that cost at the threshold will be 1, resulting in a 0% interaction probability for those located just out of it. In the case of this experiment, threshold point was arbitrarily established, although in more realistic situations, it would replicate average travel times effectively experienced in a particular city, or the invariant travel time suggested by Marchetti.

Figure 6 displays evolution of actual interaction (blue), interaction deficit (green) and M* (red) at the centre of a system growing from a 2-cell seed (position 2 in the X axe) up to 7 cells, considering Alonso population distribution, homogeneous infrastructure, and traffic effects represented by the inverse of accessibility. It is seen that M* grows with actual interaction growth from the beginning, but soon enough it falls down, crossing the X very quickly. _This performance is directly related to provision of mobility infrastructure_, in the sense that considering it evenly distributed, the system would have less than required road capacity in the central cells (and certainly more than required at the border). The green zone in the graph represents the interval in which CBD's interaction output is positive, red zone is where interaction deficit exceeds the actual one.

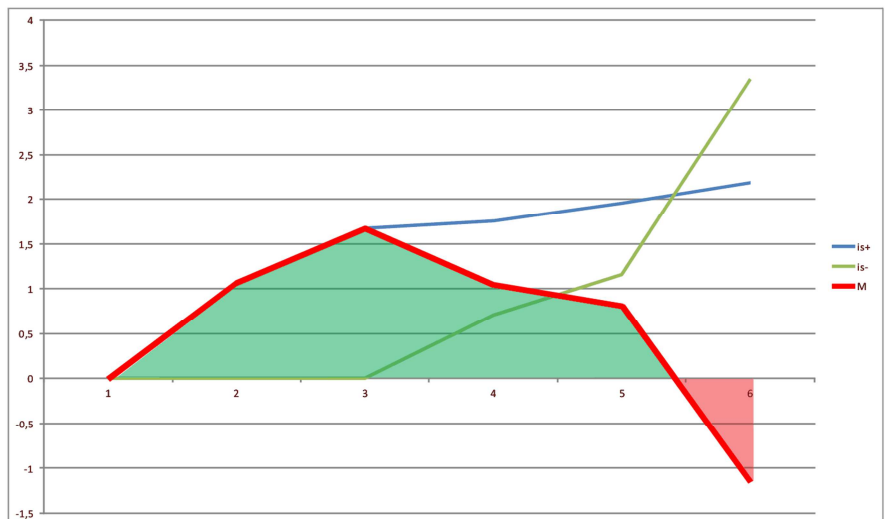

*Fig. 6: Performance indicators for the central space of a system growing from a 2-cell seed. M red curve suggests that social interaction at the centre is crucially dependent upon mobility infrastructure, as congestion effect at the most central cells cause the M indicator to fall sharply, becoming negative.*

Figure 7 displays the evolution of the same system, in which a secondary service provision location is considered at the precise cell that causes the M indicator to become negative. The introduction of a new service location introduces a competition between the two centres for potential users, which is estimated by the measure of convergence. In this particular case, convergence of the principal service centre, which had been 1, falls to 0.76 (implying that the new one starts off with 0.24 convergence. It is seen that the introduction of this new service location not only stops the M fall, but makes it to perform positively again, near the level previously experienced (but still a bit lower). The graph captures this time _the effect of spatial configuration on the social interaction_, as M performance varies with changes in service locations and consequently with relative distances and mobility infrastructure provision.

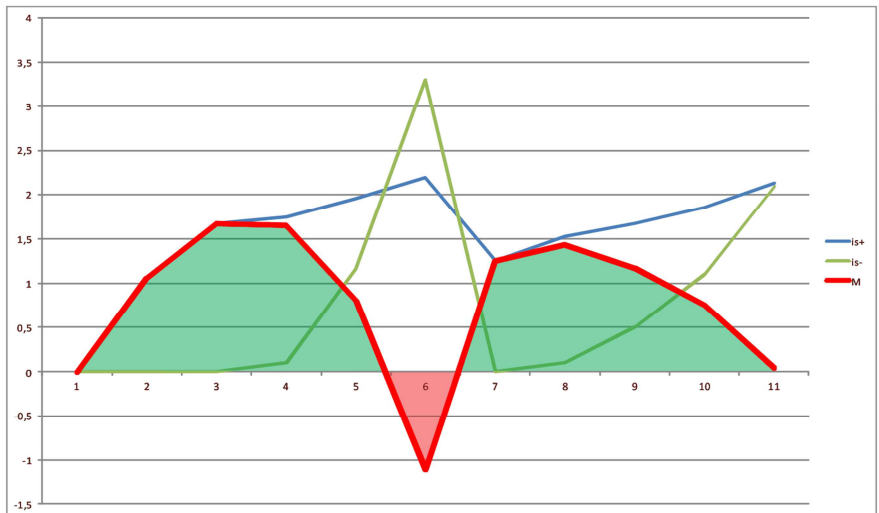

*Fig. 7: Performance indicators for the central space of a system growing from a 2-cell seed, with two service centres. M red curve suggests that social interaction at the centre is fairly dependent upon spatial configuration, as accretion of a new service location makes it return to positive performance.*

A third experiment with the same system has been carried out, this time simulating the effect of a transport system. To do this, in the last (eleventh one) step of simulation, it was assumed that, from the border, every other cell of the string were interconnected without the traffic effect (so those direct connections would be either underground or exclusive bus lanes). The display of indicators performances are in the figure 8. It is clearly noticeable the sharp upward tendency introduced in the M evolution by this, confirming the strict dependence of social interaction on mobility services. It is, then, expected that the introduction of transport in earlier stages of the system's growth would significantly change both the road infrastructure and spatial configuration's dependence of M, probably retarding and smoothing its decay.

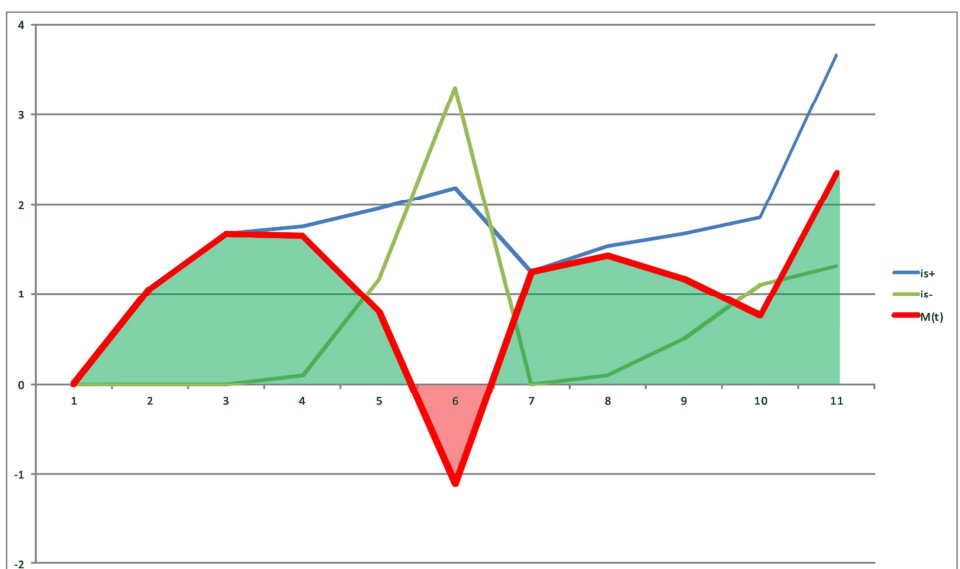

*Fig 8: Performance indicators for the central space of a system growing from a 2-cell seed, with two service centres and the introduction of a transport system in the last step of the simulation. M red curve suggests that social interaction at the centre is deeply dependent upon transportation, as accretion of a traffic-free transport line makes it shoot upwards.*

Finally, figure 9 displays, on the left side, M for all system's cells, and on the right the overall M performance. On the left, red and purple thick lines catch the two service centres, whereas the others are residential. On the right side, blue line describes the overall system's M performance without transport, and the red one, a bit displaced, shows how the transport would make things change. The first graph displays two different types of social interaction, the economically productive ones, in spaces where interaction is more intense, and the family-related ones, in residential spaces. Other types have been omitted, in spite of being real and operational. The right-side graph, summing up all the system's social interaction, combines different kinds of interaction and, in this way, should take measures to weight adequately each one. This issue has not been approached here, so that the graph is just a demonstration, focused more on the comparison of performances with and without taking transport into consideration.

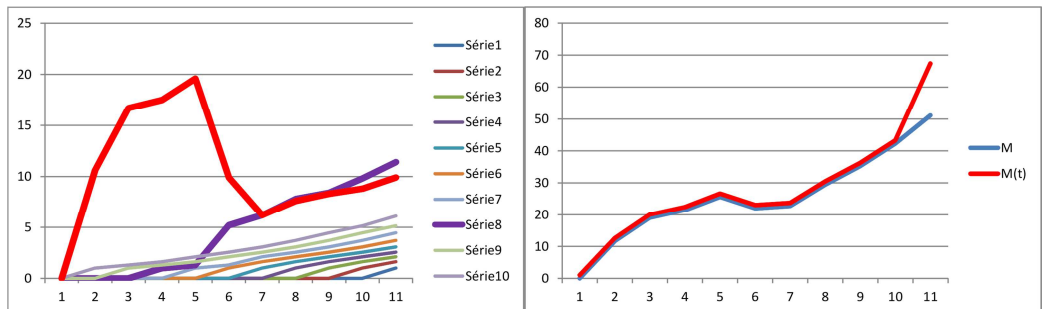

*Fig. 9: Left, M curves for all cells of the system, along evolution from 2 to 21 cells; right, M overall performance with and without transport provision.*

The simulation described above does not include interaction in the streets, in order to do that another set of calculations would be required, according the proposed model. Nevertheless, in this simple spatial configuration, it is quite predictable (increasing quantity of people in the streets from the border to the centre) and it would not change the tendencies previously described. Further levels of spatial aggregation are also left untouched; if unfolded, cell aggregation could reveal instances of social interaction potentially able to change the M overall profile, but still reinforcing the role of central spaces

FINAL REMARKS

There are two aspects of urban dynamics addressed in this work, the intertwining relationship among spatial configuration, mobility infrastructure and transportation, and the performance of spaces and parts of the city in the social interaction promotion. The model, as well as the few experiments reported in this paper, has suggested a meaningful relationship among those elements of the city, being urban form at the heart of it. Indeed, both spatial configuration and mobility infrastructure are the urban form's actual middle name, they are intrinsic to urban form; even transport systems are themselves conditioned by urban form. The reported simulation has taken urban form at its very basics, although it has allowed the necessary transparency, and so, allowed for a more controlled experimentation. It has provided early evidence on the pros and cons of the monocentric city and accumulated density, but most important, it has revealed an intricate interplay of scales within the urban system. From the most private bit of built form to the entire urban system, the city offers a variety of opportunities for social interaction, where the same individuals do perform different kinds of social interactions, within different groups. In this way, the model could have been revealing a sort of scaling functionality in the urban form, not only in its day-to-day operation but also in its long term fine tuning involving changes in location, linkage, and space provision.

The M indicator deserves a few words too; taking a relationship between effective social interaction and not fulfilled one, it could be indicating a sustainability dimension, in the sense that spaces displaying bad M scores would not be coping with its most fundamental purpose. More revealing, however, seems to be the global M, taken from the entire system, which more properly indicates the city's fulfilment of its fundamental endeavour. Nevertheless, the ability to determine each space's particular performance could be precious for planning and design purposes.

The author wishes to thank Dr. Luciana Andrade and the PROURB/UFRJ – Programa de Pós-Graduação em Urbanismo, da Universidade Federal do Rio de Janeiro, Brazil, for the support for this work.